\documentclass[pra,superscriptaddress,amsmath,amssymb,twocolumn]{revtex4-2}
\usepackage{graphicx}
\usepackage{subfigure}
\usepackage{adjustbox}
\usepackage{bm}
\usepackage[normalem]{ulem}
\usepackage{bbm}
\usepackage{color}
\usepackage{braket}
\usepackage{standalone}
\usepackage{multirow}
\usepackage{tikz}
\usepackage{mathrsfs}
\usepackage{dsfont}
\usepackage[colorlinks,bookmarks=true,citecolor=blue,linkcolor=blue,urlcolor=blue]{hyperref}
\usepackage{cleveref}
\usepackage{comment}
\usepackage{mathtools}
\usepackage{soul}
\usepackage{orcidlink}

 \Crefname{equation}{Eq.}{Eqs.}
\Crefname{figure}{Fig.}{Figs.}

\begin{document}
\title{Entropy production and statistical relaxation of dipolar bosons and fermions in interaction quench dynamics}

\author{Barnali Chakrabarti\,\orcidlink{0000-0002-6320-9894} }
\affiliation{Department of Physics, Presidency University, 86/1 College Street, Kolkata 700073, India}

\author{ N D Chavda\,\orcidlink{0000-0001-5958-1143}}
\affiliation{Department of Applied Physics, Faculty of Technology and Engineering,The Maharaja Sayajirao University of Baroda, Vadodara-390001, India.} 

\author{F.V. Prudente\,\orcidlink{0000-0002-7088-189X}}
\affiliation{Instituto de Física, Universidade Federal da Bahia, Campus Universitário de Ondina, 40170-115, Salvador, BA, Brazil}
\date{\today}

\begin{abstract}

We study the out-of-equilibrium dynamics of dipolar bosons and fermions after a sudden change in the interaction strength from zero to a finite repulsive value. We simulate the interaction quench on the initial state which is the ground state of harmonic potential with noninteracting bosons and fermions. We solve the time-dependent many-boson Schr\"odinger equation exactly using numerical methods. To understand the many-body dynamics we analyze several measures of many-body information entropy, monitoring their time evolution and assessing their dependence on interaction strength. We establish that for weak interaction quench the dynamics is statistics independent, both dipolar bosons and fermions do not relax. Whereas it is significantly different for dipolar bosons from that of dipolar fermions in the stronger interaction quench. When dipolar bosons exhibit concurrent signature of relaxation in all entropy measures, dipolar fermions fail to relax. For dipolar bosons and for larger interaction quench, the many-body information entropy measures dynamically approach the value predicted for the Gaussian orthogonal ensemble of random matrices, implying statistical relaxation. The relaxation time is uniquely determined when the orbital fragmentation exhibits a $1/M$ population in each orbital ($M$ is the number of orbitals) and all entropy measures saturate to the maximum entropy values. The relaxation time also becomes independent of the strength of dipolar interaction. Whereas, for the same quench protocol, dipolar fermions exhibit modulated oscillations in all entropy dynamics.
Our study is also complemented by the measures of delocalization in Hilbert space, clearly establishing the onset of chaos for strongly interacting dipolar bosons. It highlights the importance of many-body effects with a possible exploration in quantum simulation with ultracold atoms.
\end{abstract}
\maketitle
\section{Introduction}
Rapid experimental advances in controlling ultracold atoms and interaction have made it possible to monitor the out-of-equilibrium dynamics of the interacting Bose systems~\cite{Defenu-Andrea24,Defenu24,Defenu23}. Among the various realizations of interacting systems, one-dimensional ones are of special interest due to strong quantum correlation~\cite{Kinoshita-2004,Kinoshita-2006}. The onset of relaxation in the systems with finite number of interacting particles is an important issue in the out-of-equilibrium dynamics~\cite{Defenu24}. Statistical relaxation is manifested when some observables equilibrate in the quench process. Manipulation of highly isolated, ultracold quantum gases in low dimensions has made it possible to study the relaxation dynamics following a sudden change of parameter in the Hamiltonian. The first experiment on the relaxation dynamics of 1D atomic Bose gas has revealed a novel physics: it does not exhibit any noticeable relaxation towards equilibrium, indicating that relaxation dynamics of 1D Bose gas does not lead to thermal distribution in momentum of the atoms~\cite{Kinoshita-2006}. Such extremely slow relaxation process is indeed expected for an integrable system. Thus, the long-standing question related to the lack of thermalization has been addressed experimentally quite some time back. 

Soon after the realization of the experiment, it was established that the expectation values of few-body observables in the out-of-equilibrium dynamics of isolated interacting system can be predicted by the generalized Gibbs ensemble (GGE)~\cite{Rigol:2007,Rigol:2008}. GGEs are constructed by maximizing the entropy~\cite{JaynesI,JaynesII} and has been tested in many different systems~\cite{Rigol:2007,Rigol:2008,Rigol:2006, Thomas:2008,Martin:2008, Calabrese:2011,Calabrese:2012,Fioretto,Mossel,Calabrese:2012-2}. However, GGE fails to apply when translational invariance is broken. The meaning of relaxation in nonintegrable systems is described by the eigenstate thermalization hypothesis ETH~\cite{Rigol:2012, Srednicki} and for nearly integrable systems, higher-order statistical fluctuations are accessible experimentally~\cite{Tang:2018,Marcos:2019, Rigol:2007}. However, ETH has mainly been tested using models that employ a fixed basis set related to noninteracting particles or can be mapped to systems of hard core bosons~\cite{Rigol:2007}, fermionic Hubbard model~\cite{Kollar:2008}, the Luttinger model~\cite{Luttinger}. Calculations have been done for one-dimensional spin-1/2 systems with nearest-neighbor and next-nearest-neighbor coupling and also for gapped systems of hard core bosons~\cite{Rigol:2010,Santos:2012,Santos:2012-2}. In these statistical works statistical relaxation was manifested in the measures of Shannon information entropy and the number of principal components for strong interaction limit.

In spite of rapid experimental development, on the theoretical side, there are very few exact results (even numerically) on the non-equilibrium dynamics of interacting quantum systems~\cite{Minguzzi,Gangardt:2008,Rigol:2005,Dominik,Wilson, Yukalov,PhysRevA.109.063308,Molignini:2024,rhombik_scipost,rhombik_pre,rhombik_jpb,rhombik_quantumreports}. Furthermore, the relaxation dynamics are mostly understood for short-range interaction. However, rapid developments of experimental techniques allows controlling and manipulating atomic, molecular and optical systems~\cite{Defenu23}. Long-range quantum systems are recently realized in several experimental platforms like dipolar quantum gas~\cite{Lahaye:2009}, trapped ions~\cite{Monroe:2021,Schneider:2012}, Rydberg atoms~\cite{Saffman}, polar molecules~\cite{Carr:2009}. Long-range interacting systems also opens up avenues for potential applications in quantum technology~\cite{DeMille:2002, Rabl:2007, Weimer:2012, Islam:2013}. Ultracold atoms provide an exceptionally versatile and robust platform for realizing and tuning dipolar interactions. 
Atomic species such as erbium $^{167}$Er~\cite{Aikawa:2012}, dysprosium $^{161}$Dy~\cite{Lu:2011}, and chromium $^{53}$Cr~\cite{Griesmaier:2005}, as well as polar molecules like potassium-rubidium $^{40}$K$^{87}$Rb~\cite{Ni:2008}, sodium-lithium $^{23}$Na$^{6}$Li~\cite{Rvachov:2017}, and sodium-potassium $^{23}$Na$^{40}$K~\cite{Park:2015}, exhibit strong dipole-dipole interactions that can be precisely controlled in laboratory conditions.

Interestingly, the dynamical properties of long-range interacting systems significantly different from those of short-range interacting ones in many aspects~\cite{Schachenmayer:2013,Jens:2013,Santos:2016,Anton:2016}. For example short-range interacting systems can relax in a single time scale and long-range systems can relax in multiple time steps~\cite{Mori}. Thermalization~\cite{Rigol:2008} happens when a local observable approaches statistical equilibrium value~\cite{Peter,Anthony}. However, depending on the quench protocol, before reaching to the equilibrium value, the local observables may first settle to a quasi-stationary value leading to prethermalization~\cite{Berges,Mori:2018,Gring:2012}. In another recent work~\cite{Molignini:2024}, we establish an all-in-one out-of equilibrium dynamics for generalized long-range interacting system, demonstrating very exotic dynamics as the quench protocol was Tonks-Girardeau(TG) limit to super-TG limit~\cite{Astrakharchik:2005,Astrakharchik:2008} following Haller experiment~\cite{Haller:2009}.

In this paper, we consider a simple but realistic set up in 1D of few dipolar bosons or fermions in the ground state of a harmonic trap. In the quench dynamics, we suddenly change the interaction strength from  zero to a finite value. We study the dynamics  for the dipolar bosons by solving the time dependent Schr\"odinger equation (TDSE) numerically by employing the multiconfiguration time-dependent  Hartree method for bosons (MCTDHB)~\cite{Streltsov:2006,Streltsov:2007,Alon:2007,Alon:2008}. For the dipolar fermions we employ the fermionic variant of MCTDH, i.e., MCTDHF. Both methods are implemented in the software MCTDH-X~\cite{Alon:2008,Lode:2016,Fasshauer:2016,Lin:2020,Lode:2020,MCTDHX}. It has been established as a very efficient tool for studying the dynamics of interacting quantum gases and can evaluate the full time evolution even for strong interacting limits~\cite{Lode:2012,Cao:2013,Lin:2020-PRA,Molignini:2022,Hughes:2023}. 

Unlike the exotic dynamics, the present work likes to focus more generic case, in the quench protocol the noninteracting particles suddenly experience some repulsive interaction, ranging from weak to stronger interaction. The main objectives addressed in this work are as follows: i) To present how the dynamics is different from the exotic dynamics when the interacting particles are quenched from strongly positive to strongly attractive limit. ii) To check whether the universality of dynamics in the long-range interacting systems is a generic phenomena of interaction or the quench protocol. iii) To understand under what criteria the systems will exhibit statistical relaxation and to verify whether the relaxation process is truly dependent on quantum statistics. iv)  To provide a complete pathway of relaxation process if it happens.

To answer these questions the many-body dynamics and the relaxation process are studied by the key measures: many-body information entropy, dynamical fragmentation, delocalization in Fock space. We observe significant deviation of relaxation process as it was observed in exotic quench protocol. Our main observations are as follows: a) Quenching to weaker interaction, both dipolar bosons and fermions fail to relax, thus exhibiting universal feature in quench process when the integrable systems are weakly perturbed. b) Quenching to stronger interaction when dipolar bosons exhibit clear signature of relaxation, dipolar fermions fail to relax for the same quench protocol, thus breaking the universality of relaxation as observed both in weaker and very exotic quench protocol. c) Relaxation of dipolar bosons exhibit onset of chaos when the maximum entropy approach to the prediction Gaussian orthogonal ensemble of random matrices GOE~\cite{Kota,Izrailev:2012,Srednicki,Rigol:2008}. d) The relaxation process is inherently connected with the dynamical fragmentation and delocalization in the Hilbert space. 

The paper is structured as follows. In Sec. II, we discuss the Hamiltonian and theoretical approach. In Sec. III, we introduce quantities of interest. Section IV discusses the many-body dynamics and relaxation process explained in two subsections, for dipolar bosons and fermions. Sec. V concludes the summary.

\section{Hamiltonian and theoretical approach}

The time evolution of $N$ interacting bosons is governed by the TDSE (we set $\hbar=1$)
\begin{equation}
i \frac{\partial \Psi}{\partial t} = \hat{H} \Psi,
\end{equation}
where the total Hamiltonian for the system is
\begin{equation} 
\hat{H}(x_1, \dots, x_N)= \sum_{i=1}^{N} \hat{h}(x_i) + \sum_{i<j=1}^{N}\hat{W}(x_i - x_j)
\label{propagation_eq}
\end{equation}
The Hamiltonian $\hat{H}$ is expressed in dimensionless units, it is obtained by dividing the dimensionful Hamiltonian by $\frac{\hbar^{2}}{mL^{2}}$, with $m$ the mass of the particles and $L$ an arbitrary length scale, which we choose to be the point at which the harmonic trap equals $\frac{1}{2}$. Please refer to Appendix A for a complete discussion of how to obtain dimensionless units in MCTDH-X.
The operator $\hat{h}(x) = \hat{T}(x) + \hat{V}(x)$ is the one-body Hamiltonian containing the kinetic energy $\hat{T}(x)=-\frac{1}{2} \hat{\partial}_x^{2}$ and the external potential $\hat{V}(x)$.
The operator $\hat{W}(x_i - x_j)$ describes the two-body interaction between particles at positions $x_i$ and $x_j$. The two-body interaction can be either chosen as contact or dipolar and repulsive or attractive. For the present work we choose repulsive dipolar interaction in the interaction quench process. 

In MCTDH-X, the bosonic many-body wave function is a linear combination of time dependent permanents constructed over $M$ single-particle wave functions called orbitals.
Whereas the ansatz for the fermionic many-body wave function is equivalent, but Slater determinants replace the permanents~\cite{Lode:2016}. 
In both cases the many-body wave function $\Psi(t)$ is written as 
\begin{equation}
\left| \Psi(t) \right>= \sum_{\vec{n}}^{} C_{\vec{n}}(t)\vert \vec{n};t\rangle.
\label{many_body_wf}
\end{equation}

The vector $\vert \vec{n};t\rangle= \vert n_1,....,n_M;t\rangle; \sum_{i}n_i =N$. Thus the vector $\vec{n} = (n_1,n_2, \dots ,n_M)$ represents the occupation of each orbital, which ensures the preservation of the total number of particles. 
Distributing $N$ bosons over $M$ time dependent orbitals, the number of configurations becomes 
\begin{equation}
N_{\mathrm{conf}}^{\mathrm{boson}} =  \left(\begin{array}{c} N+M-1 \\ N \end{array}\right).
\end{equation}
\label{eq:N-conf}
For the fermionic systems, the number of configurations becomes $N_{\mathrm{conf}}^{\mathrm{fermion}} = \left(\begin{array}{c} M \\ N \end{array}\right)\approx \frac{M^N}{N!}$.

It is important to mention that the expansion coefficients $C_{\vec{n}}(t)$ and the orbitals that build up the configuration  $\vert\vec{n};t\rangle$ are time-dependent and variationally optimized at every time step ~\cite{TDVM81}. This requires the stationarity of the action with respect to variations of the time-dependent coefficients and orbitals, resulting in a coupled set of equations of motion for these quantities, which are then solved simultaneously. Note that the orbital and the coefficient $C_{\vec{n}}(t)$ are variationally optimal with respect to all parameters of the many-body Hamiltonian at any time~\cite{TDVM81,variational1,variational3,variational4}. The time-adaptive basis thus allows the sampled Hilbert space to dynamically follow the motion under any quench process. Imaginary time propagation relaxes the system to its ground state, while real-time propagation provides the full dynamics of the many-body state.

The accuracy of the algorithm depends on the number of orbitals used. For $M=1$ (a single orbital), MCTDH-X reduces to the mean-field Gross-Pitaevskii approximation. The wave function becomes exact for $M \rightarrow \infty$ as the set $ \vert n_1,n_2, \dots ,n_M \rangle$ spans the complete $N$-particle Hilbert space. For practical calculations, the Hilbert space has to be truncated by using a finite value of $M$. We repeat the simulation with gradual increase in $M$ and achieve the convergence in relevant observables, when the occupation of the highest orbital becomes insignificant. It is important to emphasize that the time-dependent orbitals assure that a given degree of accuracy is reached with a much shorter expansion compared to the time independent basis used in exact diagonalization. 

\section{Quantities of interest}
\label{sec:quantity}

In this section, we define the quantities of interest, namely the one-body density, the eigenvalues of the one-body density matrix and measures of many-body information entropy.\\

{\emph{ One-body reduced density matrix (RDM), one-body density, natural occupations and fragmentation}}.\\

Fragmentation is the hallmark of MCTDHB when more than one single particle state becomes significantly occupied. The dynamics of occupation in different orbitals offers a measure for the dynamical fragmentation in the non-equilibrium dynamics of the quantum quench.
We define fragmentation from the natural occupations $n_i$, i.e. the population of the natural orbitals, which are the eigenvalues of the reduced one-body density matrix $\rho^{(1)}(x,x';t) = \langle \psi(t) \vert \hat{\psi}^{\dagger}(x') \hat{\psi}(x) \vert \psi(t) \rangle$, i.e.
\begin{equation}
\rho^{(1)}(x, x', t) = \sum_j \frac{n_j}{N} \phi_j^*(x') \phi_j^*(x) 
\label{eq:rho1}
\end{equation}
as a spectral decomposition. In the expression for the reduced one-body density matrix (one-body RDM), $\hat{\psi}^{\dagger}(x)$ and $\hat{\psi}(x)$ are field operators for the creation and annihilation of one particle at position $x$. $\phi_j(x)$ are the natural orbitals and $n_j$ is the occupation in the respective natural orbitals.
The one-body density is determined by taking the diagonal elements of the reduced one-body density matrix, $\rho(x,t)= \rho^{(1)}(x, x'=x,t)$.
Thus for a non-fragmented or condensed state, the occupation of the lowest natural orbital is close to unity. The system becomes $k$-fold fragmented when one-body RDM has $k$ macroscopic occupied eigenvalues. Thus, for a fragmented state, several natural orbitals may contribute significantly.  Fragmentation also illustrates how close the many-body state can be described by a single wave function.
For fermions, in the non-interacting and very weakly interacting limit $M=N+1$ orbitals are sufficient satisfying exclusion principle.
However, for strongly interacting systems, a larger number of orbitals might be required to capture many-body correlations. \\

\emph{Many-body Information Entropy}\\

We calculate entropy measures such as many-body information entropy, number of principal components, occupation entropy, these are computed from the time-dependent many-body basis used in MCTDHB.
Time evolution of entropy production in the many-body dynamics are the important measures to quantify relevant physical features such as statistical relaxation and irregular or chaotic dynamics. 
To study statistical relaxation, the Shannon information entropy (SIE) $S^{\mathrm{info}}(t)$ is often considered as the ideal quantity~\cite{Berman:2004}. 
The SIE of the one-body density in coordinate space is defined as 
$S^{\mathrm{info}}(t) = -\int { \mathrm{d}x \rho(x,t) \ln \left[{\rho(x,t)} \right] }$. Analogously, for the momentum density, $S^{\mathrm{info}}(t) = -\int { \mathrm{d}x \rho(k,t) \ln \left[{\rho(k,t)} \right] }$, where $\rho(k,t)$ is the corresponding one-body density in the momentum space. However, SIE are simple measures of delocalization of the corresponding density. Since the above definitions of SIEs are related to the one-body density, they are insensitive to the correlation that may be present in the many-body state $\vert \Psi(t) \rangle$. 
Following the MCTDH ansatz (eq.(3)), we can define a few alternative entropy measures~\cite{barnali_axel, Molignini:2024} which take care of quantum correlation of the many-body state.\\

We define the many-body information entropy and inverse participation ratio in terms of time-dependent expansion coefficient of the state in the time dependent MCTDH basis. 
The modulus squared of each coefficient can be expressed as 

\begin{align}
|C_{\vec{n}}(t)|^2 = \frac{1}{\prod_{i=1}^M n_i!}& 
\left< \Psi \right| 
(\hat{b}_1(t))^{n_1} \cdots (\hat{b}_M(t))^{n_M} \nonumber \\[0.2cm]
&\quad 
(\hat{b}_1^{\dagger}(t))^{n_1} \cdots (\hat{b}_M^{\dagger}(t))^{n_M} 
\left|\Psi \right>
\end{align}
indicating that -- depending on the state -- one, some, or all $M$ creation and annihilation operators may contribute to the value of the coefficient $|C_{\vec{n}}(t)|^2$ and  provides a direct qualitative assessment of the many-body entropies within the system.

The \emph{coefficient Shannon information entropy} is defined as 
\begin{equation}
 S_{C}(t) = -\sum_{\vec{n}} | C_{\vec{n}}(t)|^2 \ln |C_{\vec{n}}(t)|^{2}.
\end{equation}
It characterizes the distribution of the state $\vert \Psi(t) \rangle$ in the Fock space.
A related quantity is the \emph{coefficient inverse participation ratio}, defined as
\begin{equation}
    I_C(t) = \frac{1} {\sum_{\vec{n}}| C_{\vec{n}}(t)|^4}.
\end{equation}
This is another measure to detect the effective number of basis states participating in the time evolution of the many-body state and is generally utilized to understand irregular dynamics. 
We also define a hybrid version of the coefficient Shannon information entropy, the \emph{$N$-body coefficient entropy} $S_C^N(t)$, as
\begin{equation}
S_{C}^N(t) = -\sum_{\vec{n}, \vec{n^{\prime}}} | C_{\vec{n}}(t)|^2 \ln |C_{\vec{n^{\prime}}}(t)|^{2}.
\end{equation}
This quantity also conveys information about delocalization in Fock space, as it inherits most of its features from the coefficient entropy $S_C(t)$, but considering all possible pairings of configurations.

Now, we will highlight the relevance of $S_C(t)$ and $I_C(t)$ in the context of mean-field theory.
In the Gross-Pitaevskii mean-field theory, only a single configuration and coefficient is included and consequently $S_C(t) =0$ and $I_C(t)=1$; $S_C(t)$ and $I_C(t)$ cannot be produced in mean-field theories. Thus $S_C(t)$ and $I_C(t)$ are the quantitative measures for how well or not a given many-body state is captured by mean-field theory.  Small $S_C(t)$, $I_C(t)$ and $S_C^{N}(t)$ mean that the many-body state contains few configurations, $|\Psi(t) \rangle$ spreads over a small part in the configuration space. The corresponding many-body state is localized. While in the quench process  the many-body state $|\Psi(t) \rangle$ may contain many configurations, the many-body state becomes dynamically delocalized, leading to larger values of $S_C(t)$, $I_C(t)$ and $S_C^{N}(t)$. When the many-body state exhibits complete delocalization, it signifies statistical relaxation, all the above three measures extracted from the time dependent coefficients will saturate to the maximum entropy state. This indicates that the MCTDHB expansion exhausted all available configuration space. This is a typical signature of strong interaction quench when the many-body states are strongly entangled. 

While these entropies provide a good grasp of the many-body character of the state and its potential progress towards a regular or irregular dynamics, they are solely based on coefficients, which are not uniquely defined and depend on the many-body state decomposition into orbitals. We thus consider another measure of entropy, \emph{occupation Shannon information entropy} and is defined as 
\begin{equation}
S_n(t)= - \sum_{i} \bar{n}_i(t) [\ln \bar{n}_i(t)],
\end{equation}
where $\bar{n}_i(t) = \frac{n_i(t)}{N}$ are natural occupations normalized with the particle number.

This entropy quantifies the distribution of particles among the various natural orbitals. 
A lower occupation entropy indicates that particles are predominantly occupying fewer orbitals (signifying non fragmented and higher coherence), whereas a higher entropy suggests a more uniform distribution across multiple orbitals (indicative of fragmentation and reduced coherence).
In a single-orbital mean-field theory, as the reduced density matrix has only a single eigenvalue, $S_n(t)=0$ and the corresponding many-body state is non-fragmented and condensed.
At the opposite extreme, when $M \to \infty$ and all the orbitals have a finite (uniform) occupation, we are in a completely chaotic regime.
In the intermediate cases, when multiple significant orbitals contribute but the occupation is not uniform at $1/M$, the corresponding many-body state is fragmented but retains a certain degree of coherence.

\section{Results}

 In this section, we present the results for quench dynamics. We perform the calculation in one spatial dimension and consider $N=4$ repulsively interacting bosons/fermions in an external harmonic trap, $\hat{V}(x_i)= \frac{1}{2}x_i^{2}$. 
 The quasi-one-dimensional dipole-dipole interaction is modeled as $\hat{W} (x_i - x_j) =g_c \delta(x_i-x_j)+\frac{g_d}{\vert x_i - x_j \vert ^{3} + \alpha_0}$. $g_c$ and $g_d$ are the coupling constants. For large separations, $r= |x_i-x_j| >> a_{\perp}$, we get the far-field long-range dipolar interaction $\sim$ $\frac{1}{r^{3}}$. For small separations, when $|x_i-x_j| \leq a_{\perp}$, the transverse confinement regularizes the divergence at $x_i=x_j$. Thus $\alpha_0 \simeq a_{\perp}^{3}$ acts as an effective interaction cutoff. The divergence is also controlled by additional $\delta$-like interaction $g_c \delta(x_i-x_j)$. The coupling constants are further  related as $g_c \simeq g_d \frac{ 4 \sqrt{\gamma}} {3 \sqrt{2\pi}}$~\cite{Santos}, where $\gamma$ is the trap aspect ratio. We refer to $g_d$ as the strength of interaction and $g_d$ could be expressed as $g_d = \frac{d_m^{2}}{4\pi\epsilon_0}$ for electric dipoles and $g_d = \frac{d_m^2\mu_0} {4\pi}$ for magnetic dipoles, $d_m$ being the dipole moment, $\epsilon_0$ is the vacuum permittivity, and $\mu_0$ is the vacuum permeability. We evaluate the cut-off parameter such that the effective interaction $V_{\mathrm{eff}}$= $\int_{\mathcal{D}} {\frac{g_d} {x^{3} + \alpha_0}} \mathrm{d} x = \int_{\mathcal{D}} { \delta(x) \mathrm{d} x} = 1 $, where the $\mathcal{D} = [-10,10]$ encompasses the entire simulation domain.  It results to $\alpha_0 \simeq 0.05$~\cite{Budhaditya}. 

The prequench state is prepared in the noninteracting limit, we first obtain the ground state for the Hamiltonian with $g_d=0$. Then at time $t=0$, we propagate the state by quenching to several $g_d$ values ranging from very weak to strong repulsion. We have probed a fine array of dipolar interactions in the range $g_d \in [0.002, 2 ]$, but we will only show representative results to be concise; $g_d=0.002,0.02,0.2$ for weak interaction quench and $g_d=1.0,1.5,2.0$ for strong interaction quench.  

To understand the quench process we perform the study with increasing orbital number $M$ and presents results for $M=10$. We present the same quench protocol for dipolar bosons and fermions. In the following two subsections, we present the quench mechanism and the entropy dynamics.

\begin{figure}
\centering
\includegraphics[width=1.0\columnwidth]{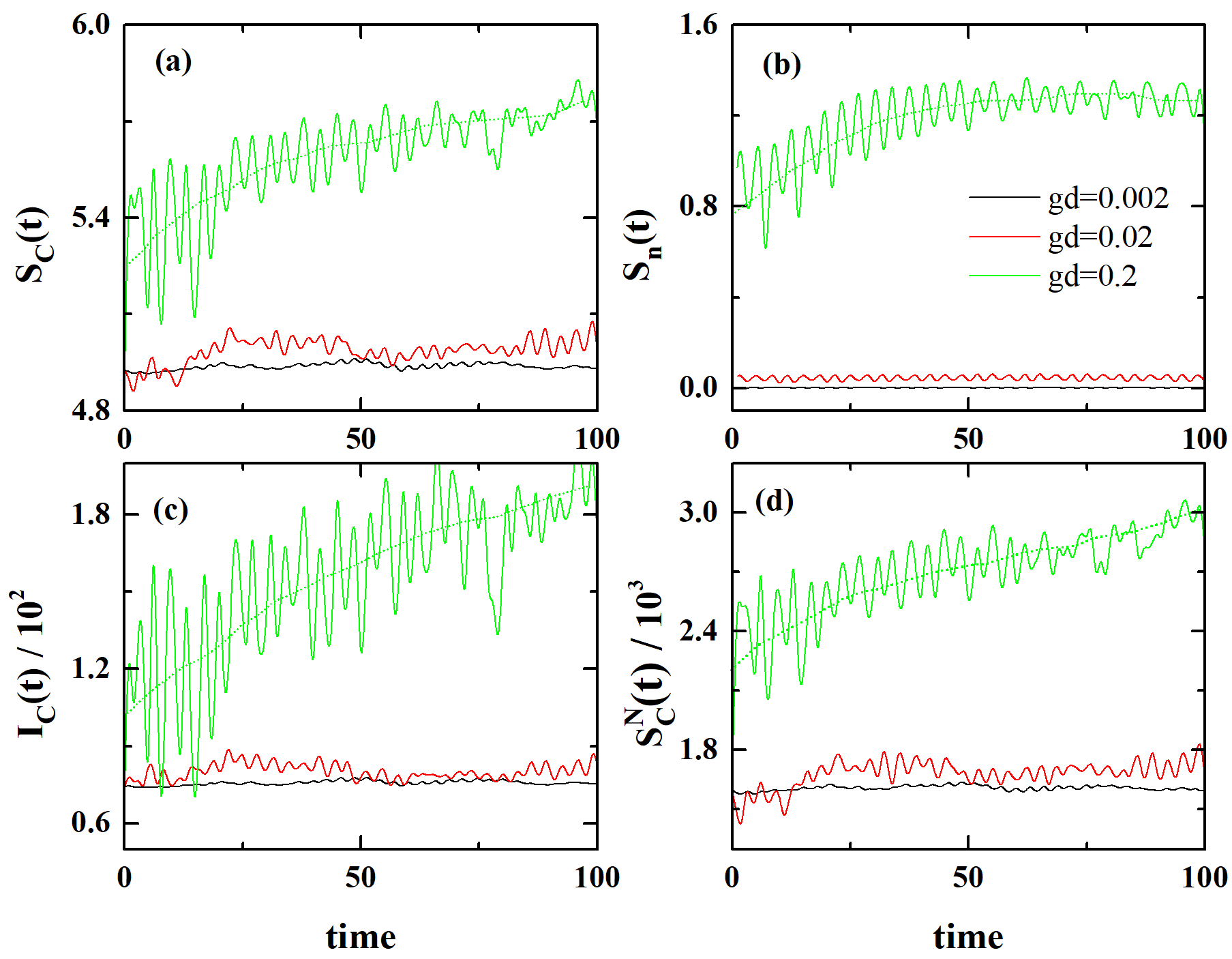}
\caption{Time evolution of four different entropy measures for $N=4$ quenched dipolar bosons in the weak interaction limit, $g_d=0.002,0.02,0.2$. a) coefficient entropy $S_C(t)$, b) occupation entropy $S_n(t)$, c) coefficient inverse participation ratio $I_C(t)$, d) many-body coefficient entropy $S_C^{N}(t)$. The overall observation is same for $g_d=0.002$ and $0.02$, i.e, the many-body state can be described in mean-field theory. For $g_d=0.2$, large fluctuations emerge in all quantities due to absence of strong correlations between the particles. The thin dashed lines are provided to guide the eye and to estimate the magnitude of fluctuations. See the text for details.  }

\label{fig1}
\end{figure}

\subsection{Relaxation process and Entropy dynamics for dipolar bosons}

\begin{figure}
\centering
\includegraphics[width=1.0\columnwidth]{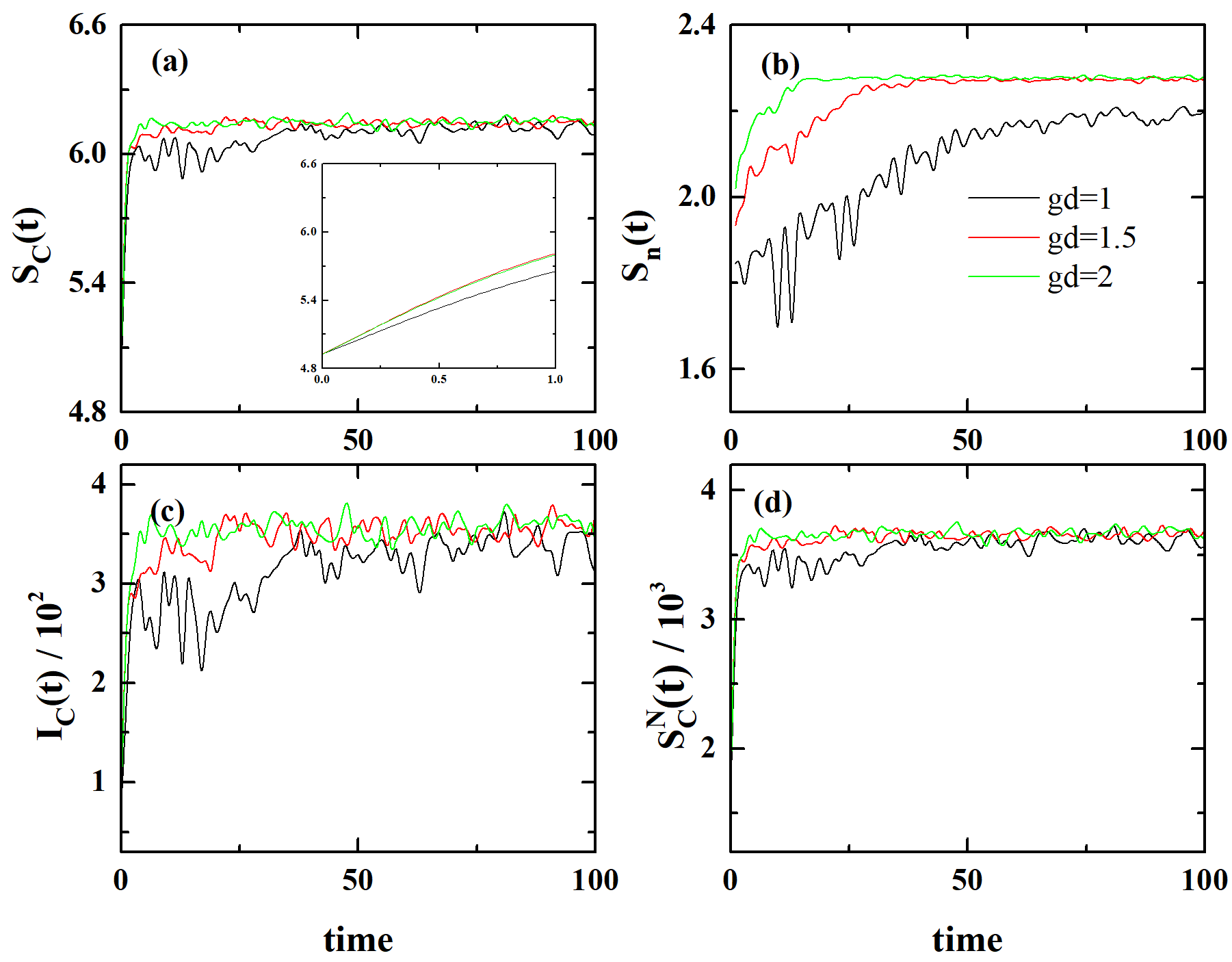}
\caption{Time evolution of four different entropy measures for $N=4$ quenched dipolar bosons in the strong interaction limit, $g_d=1.0, 1.5, 2.0$. a) coefficient entropy $S_C(t)$, the inset exhibits linear increase in entropy in short time, b) occupation entropy $S_n(t)$, c) coefficient inverse participation ratio $I_C(t)$, d) many-body coefficient entropy $S_C^{N}(t)$. Statistical relaxation happens as all the entropy measures manifest the convergence to GOE predictions. See the text for details.}

\label{fig2}
\end{figure}

We begin by probing the relaxation process of dipolar bosons considering a sudden quench from $g_d=0.0$ to finite $g_d$ values. Fig.~\ref{fig1} summarizes measures of all the four entropic quantities $S_C(t)$, $S_n(t)$, $I_C(t)$ and $S_C^{N}(t)$ for weak interaction quench whereas Fig.~\ref{fig2} discusses the same measures for strong interaction quench. 

The key aspect of many-body quantum chaos is how the many-body wavefunction spreads across the available basis functions with an increase in the interparticle interaction strength. The spreading and the fragmentation are counteracted by the time adaptivity of the basis set in MCTDHB. For an assessment of the presence of statistical relaxation, we visualize the time evolution of the different entropy measures, delocalization in Hilbert space and dynamical fragmentation. 

First, we discuss the case for weak interaction quench with $g_d$ = 0.002, 0.02 and 0.2. All entropy measures are plotted in Fig.~\ref{fig1}.
For $g_d$=0.002 and 0.02, we observe that all four measures $S_C(t)$, $S_n(t)$, $I_C(t)$ and $S_C^{N}(t)$ remain constant throughout the dynamics, signifying that the dipolar bosons do not respond to such weak interaction quench. We will discuss later how the constant value in $S_C(t)$ is related with the dynamical localization. Also $S_n(t)$ is close to zero, signifying the many-body state can be defined in the mean-field level. For slightly stronger interaction quench, $g_d=0.2$, when the interatomic correlations do not play much significant role, all the measures exhibit large fluctuations ruling out any possibility in the emergence of a saturation. 

The results for strong interaction quench, $g_d$= 1.0, 1.5 and 2.0 are presented in Fig.~\ref{fig2}. We observe a concurring estimate from all the four different measures that the relaxation happens around $t \approx 20$, when all the measures equilibrate. However, for interacting many-body systems, statistical relaxation is also related to chaos in the energy spectra. The onset of chaos is established when the entropy measures first exhibit a quadratic growth at the very short time, then linear growth before saturation to the maximum entropy state. For time reversal and rotationally invariant systems, the maximum entropy is also determined by the predictions of a Gaussian orthogonal ensemble of random matrices(GOE)~\cite{Kota,Rigol:2008,Izrailev:2012,Srednicki}. We present how the different measures approach the GOE prediction as determined by the number of orbitals used in the computation. 
For a GOE of random matrices,  $S_C^{GOE}=\ln 0.48 D$ and $I_C^{GOE}=D/3$, where $D\times D$ is the dimension of the random matrices~\cite{Kota}.
To compare with GOE estimates, we set $D$ as equal to the number of configurations $N_{\mathrm{conf}}^{\mathrm{boson}}$ participating in the dynamics. For occupation entropy $S_n$, $D$ is set equal to the number of orbitals $M$ used in the computations. 
Thus $S_n^{GOE}$ = $-\sum_{i=1}^{M} \frac{1}{M} \ln(\frac{1}{M}) = \ln(M)$. 

For the present calculation with $M=10$, the number of configurations used in the simulation is $N_{\mathrm{conf}}^{\mathrm{boson}}=715$. The corresponding GOE predictions are $S_C^{\mathrm{GOE}} \approx 5.84$, $S_n^{\mathrm{GOE}} \approx 2.30$, and $I_C^{\mathrm{GOE}}= 238$. 
The saturation value from numerical analysis are  $S_C^{\mathrm{sat}} \approx 6.13$, $S_n^{\mathrm{sat}} \approx 2.27$ and $I_C^{\mathrm{sat}}= 350$. These values are indeed close to GOE prediction, however small discrepancy exists. The perquisite for GOE prediction is the infinitely strong interaction, for our numerical simulation the interaction quench is strongly positive but not infinite. The initial linear growth followed by the saturation close to GOE prediction indeed exhibits the hallmark of statistical relaxation. Fig.~\ref{fig2} also demonstrates that once relaxation happens, both the relaxation time and the maximum entropy values remain unchanged with further increase in $g_d$ values. Next, we will exhibit that statistical relaxation happens when complete delocalization happens in the Hilbert space signifying the onset of chaos.

\begin{figure}
\centering
\includegraphics[width=1.0\columnwidth]{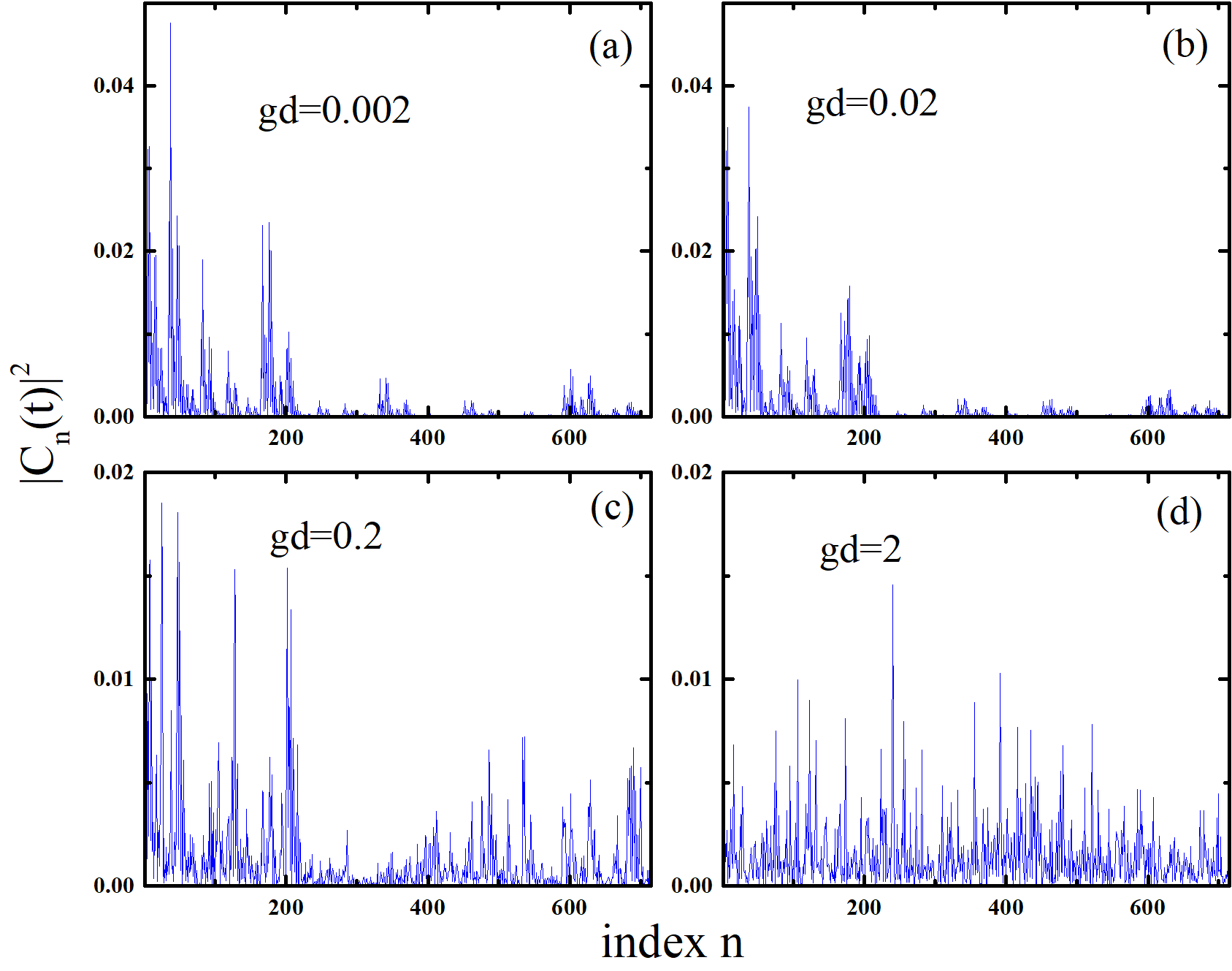}
\caption{Configuration space population at time $t=100$, the distribution of the magnitude of the coefficients $\{ |C_{\vec{n}}(t)|^{2}\}$ as a function of index $n$ for different interaction quench. The index $n$ is computed from vector $\vec{n}$ using the mapping described in Ref.~\cite{Streltsov:2010}. Figs (a)-(d) represent the magnitude of the coefficients for (a) $g_d=0.002$, (b) $g_d=0.02$, (c) $g_d=0.2$, (d) $g_d=2$. The state stays rather localized for $g_d=0.002$ and $0.02$. With increasing in the interaction strength $g_d$, more coefficients in the expansion become significant. The state rapidly becomes delocalized.  }

\label{fig3}
\end{figure}

Understanding the relaxation process is an important challenge in the interacting quantum many-body dynamics. To assess the relaxation it is instructive to study delocalization in the Hilbert space and dynamical fragmentation. In Fig.~\ref{fig3}, we plot $|C_{\vec{n}}(t)|^2$ as a function of the index $n$ of the basis states for the final simulation time at $t=100$ for different interaction quench, $g_d$=0.002, 0.02, 0.2, 2. We refer reference\cite{Streltsov:2010}, for the details on how to calculate the index $n$ from the vectors $\vec{n} = (n_1,...n_M)$. Note that the permanents $ \vert \vec{n};t \rangle$ are not eigenstates of the Hamiltonian for  $t>0$. However, every eigenstate of the interacting system can be represented as a pattern of coefficients that contribute significantly. 
The number of configurations in the available space for particular choice of $M=10$ is $N_{\mathrm{conf}}^{\mathrm{boson}}$ =715. As depicted in Figs.~\ref{fig3} (a)-(d), the number of significantly contributing coefficients grows with the interparticle interaction strength. For relatively weak interaction, $g_d=0.002$ and $0.02$, the number of significantly contributing coefficients is only a small portion of the total number of configurations. We find, out of available $715$ configurations, only $200$ configurations participate significantly in the dynamics leading to the constant measure of coefficient entropy $S_C=4.94$ and $I_C= 70$ as shown in Fig.~\ref{fig1}(a) and (c). We refer to such a state as localized due to absence of interatomic correlation. We emphasize that the term "localized" refers to many-boson Fock space and not real space. However, for $g_d=0.2$, we monitor the time evolution of the distribution of the magnitude of the coefficients  in the Hilbert space, with increasing time, more coefficients in the expansion become significant (not shown here). At the final time of simulation $t=100$, the contributing coefficients spanned by the entire configurations, but the coefficients are not uniformly distributed all over the index, leading to modulation in entropy measures as shown in Fig.~\ref{fig1}(a),(c) and (d). For $g_d=2$, the coefficients explore almost the whole available space spanned by $N_{\mathrm{conf}}^{\mathrm{boson}}=715$, the complete delocalization happens leading to statistical relaxation as shown in Fig.~\ref{fig2}. Thus statistical relaxation is characterized by delocalization of many-body state in the Hilbert space manifesting the onset of chaos.

\begin{figure}
\centering
\includegraphics[width=1.0\columnwidth]{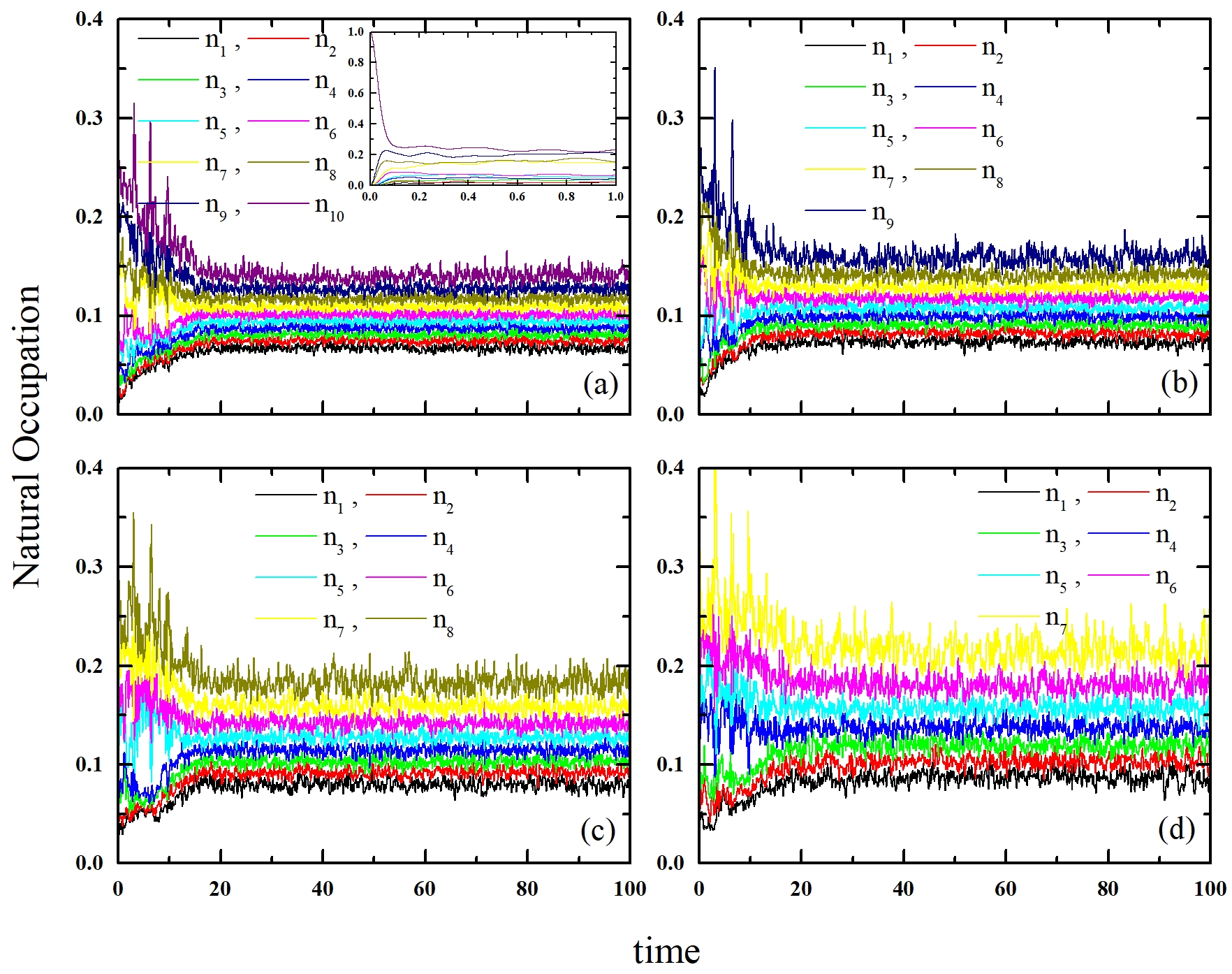}
\caption{Orbital fragmentation dynamics for $N=4$ interacting dipolar bosons quenched to $g_d=2$ and computation is done with different orbitals. (a), $M=10$, in the inset, population in ten natural orbitals up to time $t=1$, (b) $M=9$, (c) $M=8$, (d) $M=7$. For all cases approximately at time $t \approx 20$, the population of the orbitals coalesces around the value $1/M$. See the text for details.}
\label{fig4}
\end{figure}

Fig.~\ref{fig4} depicts the orbital occupation as a function of time when the noninteracting dipolar bosons are quenched to $g_d=2$. Fig.~\ref{fig4}(a), shows the orbital fragmentation for $M=10$ orbitals, which are also used in the computation of entropy dynamics throughout the work. The plot shows that initially the system is fragmented, leading to nonzero value of $S_n(t)$ (Fig.~\ref{fig2}(b)). The inset exhibits that, the occupation in the significantly populated orbitals decreases with time while the initially nearly empty orbitals start to populate. Approximately at time $t \approx 20$, the population of all ten orbitals coalesces around the value $\frac{1}{M}=0.1$. This irregular behavior termed as violent fragmentation  and is also observed in the exotic quench process in our earlier work~\cite{Molignini:2024}.  
Before we remark the violent fragmentation is the generic feature of strong dipolar interaction quench, we repeat all the simulations with $M=9, 8, 7$ and the corresponding orbital fragmentation is presented in Fig.~\ref{fig4}(b)-(d) respectively. It is observed that at the transition time, $t \approx 20$ when the dipolar bosons exhibit relaxation, all the available orbitals become populated with $1/M$ occupation in each orbital. 

We conclude the following four facts: i) Strongly interacting dipolar bosons exhibit statistical relaxation in the out-of-equilibrium dynamics. ii) The relaxation happens when the many-body state exhibits {\emph{complete delocalization and exotic orbital fragmentation}}. iii) The time of relaxation is determined by concurring estimate from {\emph{saturation of all entropy measures, time required for delocalization and complete fragmentation}}. iv) Weakly interaction quench does not lead any relaxation.

\subsection{Relaxation process and Entropy dynamics for dipolar fermions}

\begin{figure}
\centering
\includegraphics[width=1.0\columnwidth]{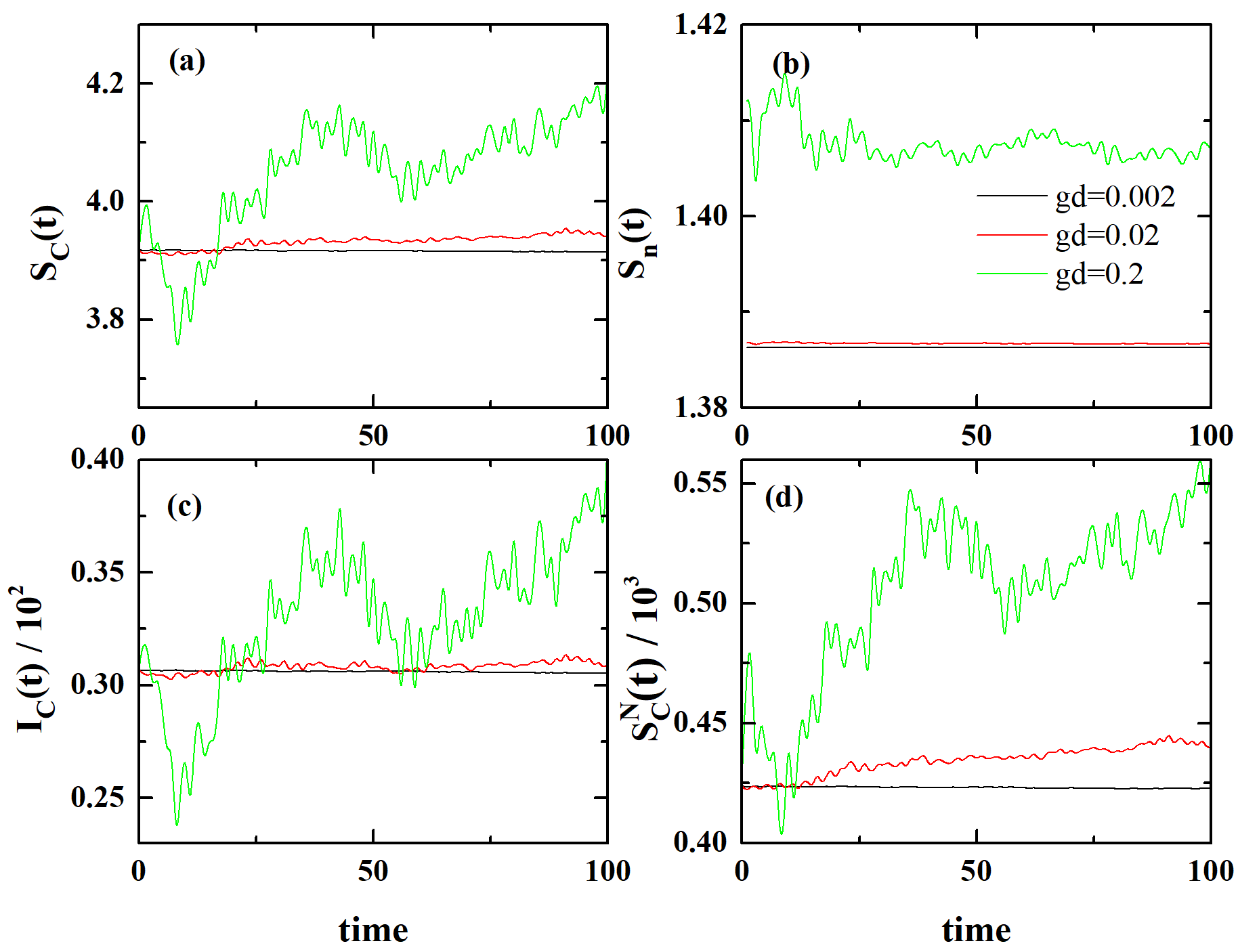}
\caption{Time evolution of four different entropy measures for $N=4$ quenched dipolar fermions in the weak interaction limit, $g_d=0.002,0.02,0.2$. a) coefficient entropy $S_C(t)$, b) occupation entropy $S_n(t)$, c) coefficient inverse participation ratio $I_C(t)$, d) many-body coefficient entropy $S_C^{N}(t)$. Possibility of relxation is ruled out. See the text for details.}

\label{fig5}
\end{figure}

\begin{figure}
\centering
\includegraphics[width=1.0\columnwidth]{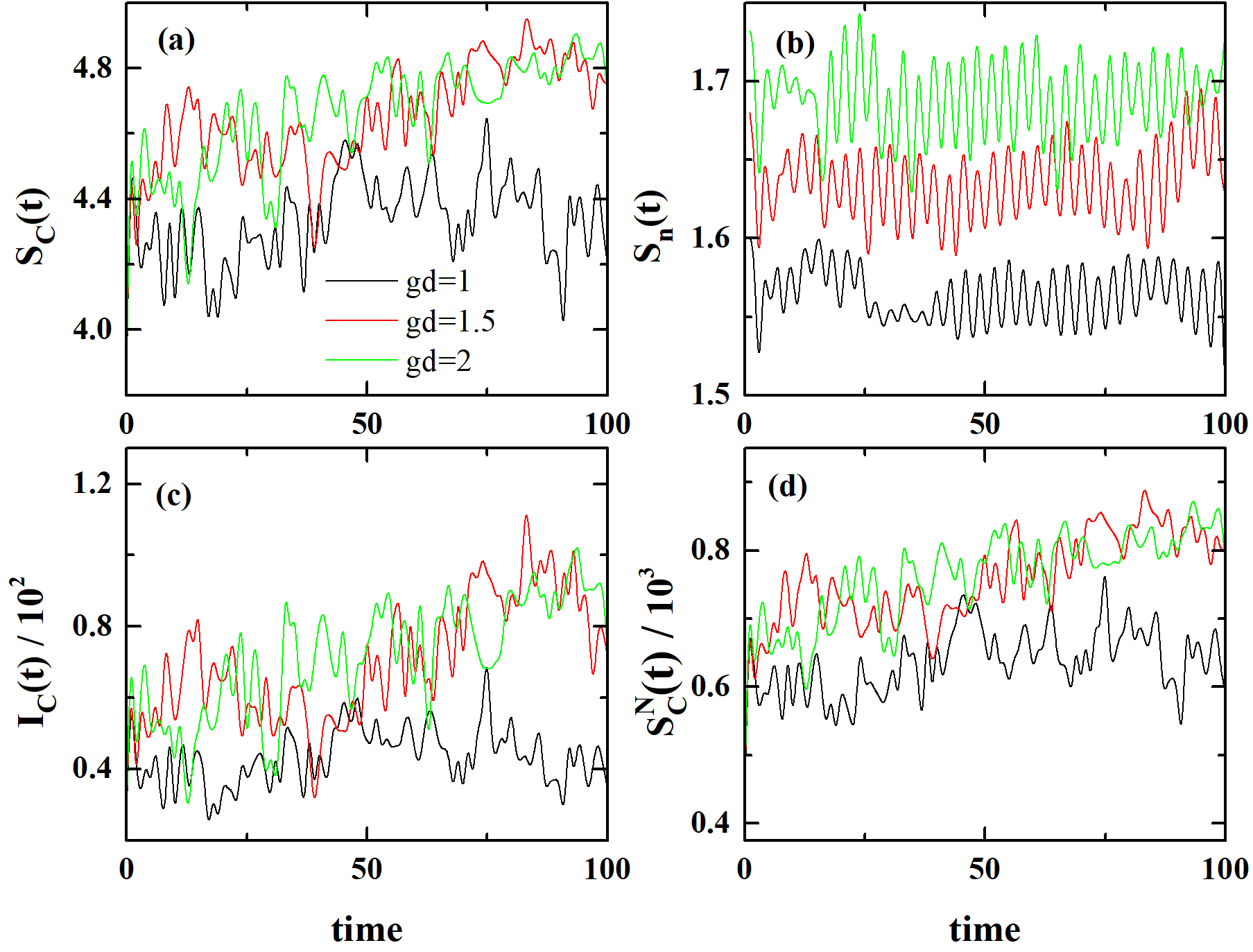}
\caption{Time evolution of four different entropy measures for $N=4$ quenched dipolar fermion in the strong interaction limit, $g_d=1.0, 1.5, 2.0$. a) coefficient entropy $S_C(t)$, b) occupation entropy $S_n(t)$, c) coefficient inverse participation ratio $I_C(t)$, d) many-body coefficient entropy $S_C^{N}(t)$. Unlike the dipolar bosons, dipolar fermions do not undergo any relaxation process. See the text for details.}

\label{fig6}
\end{figure}

In this section, we present results for many-body dynamics for $N=4$ spinless \emph{dipolar fermions}, maintaining the same prequench and postquench protocol chosen for the bosonic system. 
The corresponding measures of entropy are plotted in Fig.~\ref{fig5} for weaker quench, $g_d$ = 0.002, 0.02, 0.2 and in Fig.~\ref{fig6} for stronger quench 
$g_d$ = 1.0, 1.5, 2.0. We observe that the relaxation process strongly depends on quantum statistics. 

For very weak interactions, $g_d$ =0.002 and 0.02, all the entropy measures are almost constant in the entire time evolution and the constant values can be further related with the delocalization in Hilbert space and dynamical orbital population discussed later. For slightly stronger interaction, $g_d$=0.2, we observe highly modulated oscillatory structures in the dynamics of $S_C(t)$, $I_C(t)$ and $S_C^N(t)$. 
This makes it hard to determine the exact nature of dynamics. 
However, we find that the occupation entropy exhibits a smoother behavior, although there is no signature of statistical relaxation for which the entropy should exhibit some linear increase before reaching some saturation value. Instead, Fig.~\ref{fig5}(b) shows fluctuating oscillation up to $t \simeq 30$ and then exhibits smooth oscillation. We will discuss that dynamical features manifested in Fig.~\ref{fig5} can be understood if we can evaluate how much Hilbert space really participate in the dynamics and also how many orbitals become significantly occupied in the dynamical evolution.

The results for higher interaction quenches are plotted in Fig.~\ref{fig6}. $S_C(t)$ exhibits oscillatory structure in the entire dynamics, ruling out of statistical relaxation even for longer time dynamics or with much stronger interaction strength. Later we will establish that the many-body system with few dipolar fermions remain localized in the Hilbert space and the fluctuating behavior can be manifested when the occupation in the Hilbert space fluctuates. We conclude the same physics for $I_C(t)$ and $S_C^{N}(t)$. The occupational entropy dynamics exhibits fluctuation about some constant value. The oscillatory pattern will be further connected by the oscillation in dynamical orbital population when the several natural orbitals populate and depopulate in a fixed interval of time. Also the average value smoothly increases with increase in $g_d$ values, that can also be explained how the higher orbitals start to fill up gradually with increases in $g_d$ values as discussed later. Thus the observed dynamics in Fig.~\ref{fig6}(b) is simply the manifestation of orbital population, $S_n(t)$ does not exhibit any generic signature of relaxation process, it neither shows any linear growth at very short time nor saturation to equilibrium value determined by the GOE prediction. All the entropy measures concurrently conclude the same physics that the dipolar fermions do not relax in the interaction quench dynamics. It may exhibit relaxation only for exotic quench process as discussed in our earlier work~\cite{Molignini:2024}.

\begin{figure}
\centering
\includegraphics[width=1.0\columnwidth]{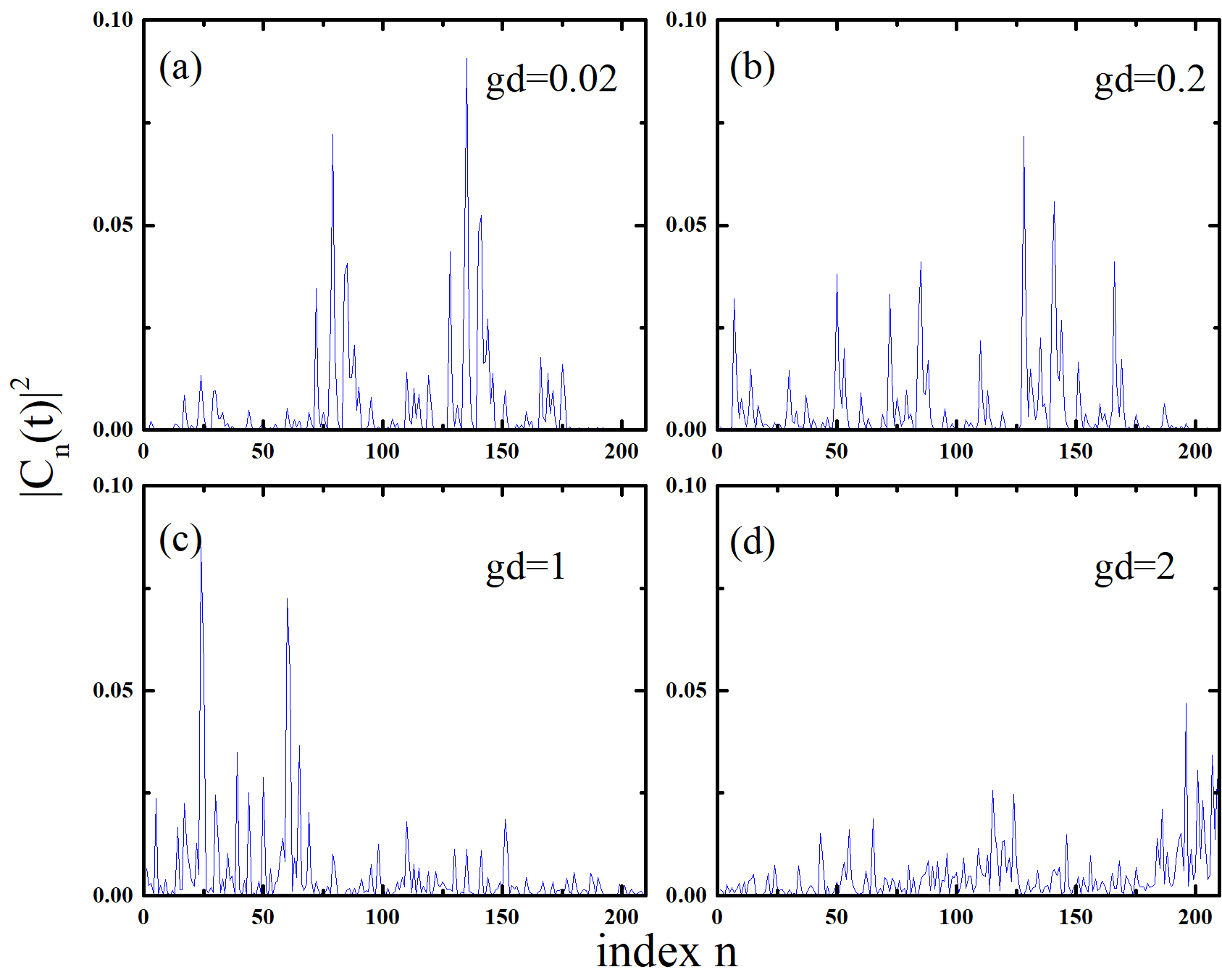}
\caption{Configuration space population at time $t=100$ for $N=4$ dipolar fermions quench dynamics, the distribution of the magnitude of the coefficients $\{ |C_{\vec{n}}(t)|^{2}\}$ as a function of index $n$ for different interaction quench.
The index $n$ is computed from vector $\vec{n}$ using the mapping described in Ref.~\cite{Streltsov:2010}. Figs (a)-(d) represent the magnitude of the coefficients for (a) $g_d=0.002$, (b) $g_d=0.02$, (c) $g_d=0.2$, (d) $g_d=2.0$. The spreading of coefficients almost remain invariant on the values of $g_d$ and stays rather localized ruling out any possibility of relaxation.  }

\label{fig7}
\end{figure}

We conclude that out-of-equilibrium dynamics of dipolar fermions is distinctly different from that of dipolar fermions when they are driven far from equilibrium by  the same quench protocol. 

To further understand the observed dynamics in Figs.~\ref{fig5} and~\ref{fig6}, we additionally analyze the population in Hilbert space and dynamical population. In Fig.~\ref{fig7} we plot the expansion coefficients $|C_{\bar{n}}(t)|^2$ as a function of the index $n$ of the basis states for the final simulation time at $t=100.0$ for different interaction quench, $g_d=0.02, 0.2, 1.0, 2.0$. For the fermionic system the number of configurations used in the simulation is $N_{\mathrm{conf}}^{\mathrm{fermion}} \approx \frac{M^N}{N!}$. Thus the present computation with $N=4$ dipolar fermions and $M=10$ orbitals results to $N_{\mathrm{conf}}^{\mathrm{fermion}}=425$. The striking observation is even for the strongest interaction $g_d=2.0$, less than $50\%$ of the available Hilbert space is occupied. It signifies that even in the strongest interaction quench, the system of dipolar fermions does not exhibit delocalization in the Hilbert space which is inherently related with statistical relaxation, concluding the fact that unlike the dipolar bosons, dipolar fermions do not exhibit relaxation. 

\begin{figure}

\includegraphics[width=1.0\columnwidth]{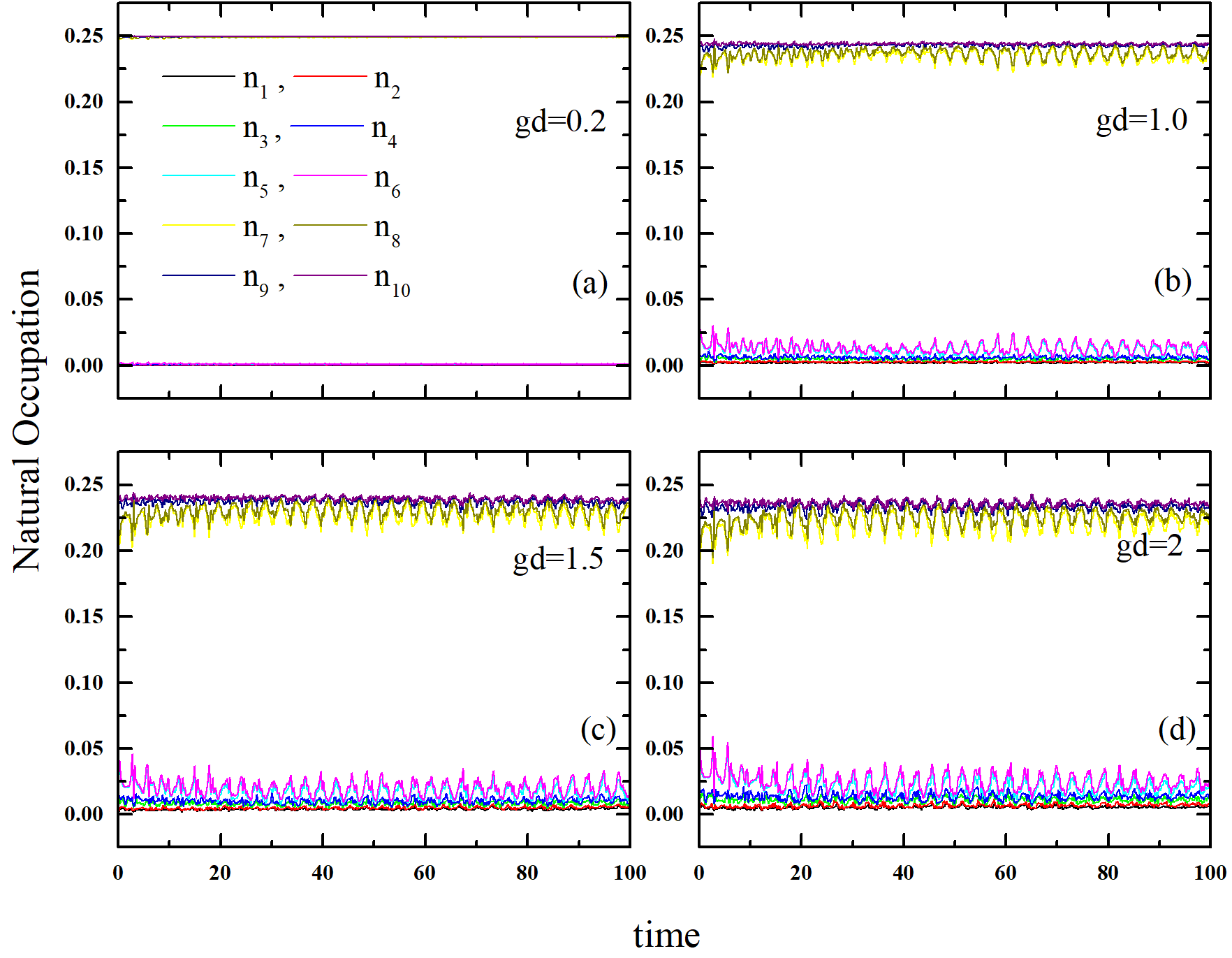}
\caption{Orbital population dynamics for $N=4$ interacting dipolar fermions quenched to different $g_d$ values. (a), $g_d=0.2$, the lowest $M=N=4$ orbitals contribute of weight $1/M= 25 \%$, when all the rest orbitals are empty. (b) $g_d=1.0$, the lowest four orbitals contribute $\approx 23 \%$, whereas few other orbitals contribute the rest $8\%$. For (c) $g_d=1.5$ and  (d) $g_d=2.0$, the contribution from the lowest four orbitals smoothly decrease whereas other orbital contribution increases slowly. See the text for details.}

\label{fig8}
\end{figure}

In Fig.~\ref{fig8} we plot dynamical population in several orbitals for different interaction quench. The occupations in all $M=10$ natural orbitals are plotted as a function of time. 
For weaker quench process with $g_d=0.2$, it is uniquely established that the number of dominating orbitals is clearly the number of fermions exhibiting the Pauli principle, each fermion resides in a single orbital. For $N = 4$, the
first four natural orbitals contribute equally $\approx 25 \% $, when the rest orbital occupation is essentially zero. With the $N=4$ contributing orbitals, the expected occupational entropy should be $S_n= ln(M=4)= 1.38$ which is close to the constant value presented in Fig.~\ref{fig5}(b). With increasing values of $g_d$, the major contribution comes from the lowest four orbitals, however few other empty orbitals start to contribute, leading to variation in $S_n$ between $1.61$ and $1.78$, these are already manifested in Fig.~\ref{fig6}(b). Thus the average value of fluctuations in the occupation entropy can be estimated via orbital population dynamics. However dynamics of $S_n(t)$ does not exhibit any signature of relaxation, oscillating about an average value in the entire dynamics. However, the typical unmodulated oscillation in $S_n(t)$ is also visualized in the dynamics of orbitals population in the same time scale.

We conclude the following facts. i) Weakly interacting dipolar fermions and bosons exhibit the same features in out-of-equilibrium dynamics, i.e., the failure to achieve relaxation.
ii) In the stronger interaction quench, the dynamics of dipolar fermions are significantly different from dipolar bosons. Even in the strongest interaction quench studied in the present work, dipolar fermions do not manifest any signature of relaxation. iii) The fluctuating dynamics in $S_C(t)$ is manifested by Hilbert space population, however unlike dipolar bosons, dipolar fermions do not exhibit complete delocalization, signaling no possibility of statistical relaxation or onset of chaos. iv) The fluctuating dynamics in $S_n(t)$ is characterized by the orbital population dynamics, however $S_n(t)$ does not exhibit the generic signature of statistical relaxation.

\section{Conclusion}

Understanding out-of-equilibrium dynamics is one of the major unsolved  problems in quantum mechanics. The scarcity of appropriate theoretical tool has made this research area more challenging. Although it is an established fact that a generic isolated quantum system relaxes, the process of relaxation is not known. Relaxation may happen in a short or long time scale and can be highly complex. Nearly integrable quantum systems, which fail to relax even in the long time dynamics, can exhibit an intermediate quasi-equilibrium state, known as prethermal state, although the origin of prethermal states is still unknown. 

The introduction of long-range interaction can play a key role and the process of out-of-equilibrium dynamics may be more fascinating. Ultracold dipolar bosons and fermions would be an ideal platform to study interesting features in out-of-equilibrium dynamics. In the present work, we have studied relaxation process of parabolically trapped interacting dipolar bosons and fermions for an interaction quench. We have solved the full time-dependent solution of the many-body problem with MCTDH-X for the continuum system. The time dependent variational optimization and decomposition of the many-body wavefunction into an adaptive basis set of time-dependent single particle orbitals has made the entire simulation numerically exactly. 

We have studied the entire relaxation process by measuring many-body entropy measures and establish a connection that an increase in the many-body entropy measures and equilibration to the value predicted by GOE estimate is the hallmark of statistical relaxation. It is important to emphasize that the saturation is not a trivial phenomenon originated from the fact that optimization of the basis set in MCTDHB that minimizes the portion of significant coefficients in the expansion. The coefficient and occupational entropy gradually approach the GOE values dynamically and statistical relaxation prevails for the considered Hilbert space.  

Overall, we have observed that for weak interaction quench, both dipolar fermions and bosons fail to relax to the maximum entropy state as predicted by GOE. Rather, all the entropy measures either remain constant or strongly fluctuate manifesting the lack of many-body correlation. Out-of-equilibrium dynamics of dipolar fermions are in strong contrast as that of dipolar bosons for stronger interaction quench. For dipolar bosons, statistical relaxation is established in a single time scale with clear signature of onset of chaos. Additionally, we establish that relaxation is also associated with delocalization in Hilbert space and orbital fragmentation. The relaxed state is characterized by the maximum entropy state. Whereas, dipolar fermions fail to relax even in the strongest interaction quench reported in the present simulation. All the entropy measures exhibit modulated oscillation in the entire dynamics. Therefore the comprehensive analysis of the quench protocol establish the non-universal features in the relaxation process of dipolar fermions and bosons. Thus the out-of-equilibrium dynamics is statistically dependent. They become universal only for the exotic quench protocol. 

The theoretical findings obtained in the work, can be tested in various experimental set ups of dipolar systems. The immediate extension of the present work is to explore the dynamics of non-exotic quench protocol address in the present work but for the generalized long range interaction $\frac{1}{r^{\alpha}}$, where $\alpha$ is the decay constant.

\section*{Acknowledgment}
We acknowledge the partial financial support provided by the Brazilian “Coordenação de Aperfeiçoamento de Pessoal de Nível Superior” (CAPES)—Finance Code 001.

\appendix

\section{Units}

In this appendix, we discuss the units for the simulations presented in the main text.
The system consists of $N=4$ bosons or fermions in an optical harmonic trap of frequency $\omega$.

The units of our simulations are chosen as follows.
We choose to set the unit of length as the inverse of the harmonic trapping frequency, i.e. $\bar{L} \equiv \sqrt{\hbar/(m\omega)}$.
We then run simulations with 512 grid points in an interval $x \in [-10 \bar{L}, 10 \bar{L}]$, giving a resolution of 0.039$\bar{L}$.

The unit of energy $\bar{E}$ in MCTDH-X is defined in terms of the unit of length as $\bar{E} \equiv \frac{\hbar^2}{m \bar{L}^2}$. By inserting our choice for the unit of length, we immediately see that the unit of energy corresponds to the quantized energy of the harmonic trap, i.e. $\bar{E} = \hbar \omega$.
During the quench procedure, we change the strength of the long-range interactions from an initial value of $g_d=0 \bar{E}$ to a positive value between $g_d=0.002 \bar{E}$ and $g_d=2.0 \bar{E}$, as explained in the main text, ranging from weak to strong interaction.

The unit of time in MCTDH-X is also defined from the unit of length, as 
$\bar{t} \equiv \frac{m \hat{L}^2}{\hbar}$.
Again inserting our definition of unit of length, we can simplify $\bar{t}=\frac{1}{\omega}$, i.e. the unit of time is the inverse frequency of the trap. 
In our simulations, we ran most time evolutions up until around $t \approx 1.2 \bar{t}$ as this time scale is enough to probe the relaxation dynamics of the quench.

\bibliography{biblio}

\end{document}